# AnatomIQ: An Open-Source Toolkit for Automated Background Detection in Medical Imaging

Rafael Carballeira[1], Hayley A. Cash[3], Marthony L. Robins[1,2]

[1]Thayer School of Engineering, Dartmouth College, Hanover, NH

[2]Dartmouth Hitchcock Medical Center, Lebanon, NH

[3]Medical University of South Carolina, Charleston, SC

## Abstract

Manual background selection for contrast-to-noise ratio (CNR) calculations in CT image quality assessment is time-consuming, operator-dependent, and compromises reproducibility. Advanced metrics such as Noise Power Spectrum (NPS) and Task Transfer Function (TTF) traditionally require dedicated phantom acquisitions, adding cost, radiation exposure, and workflow disruption. We developed AnatomIQ, an open-source Python toolkit that automates background detection and performs phantom-free NPS and TTF analysis directly on patient anatomy. Automated background detection uses tissue-specific Hounsfield unit thresholding and morphological filtering, paired with a web interface for click-to-select lesion identification from PACS screenshots. Reported metrics include CNR, signal-to-noise ratio, detectability indices, and dose-normalized figures of merit with automated optimization recommendations. NPS quantifies noise texture from automatically detected uniform tissue regions, while TTF leverages natural circular anatomical structures (e.g., trachea, vessels) as edge phantoms, together characterizing the resolution-noise trade-off without dedicated phantom scans. Analysis takes 2-3 minutes per case, with CNR computed in under 10 seconds and NPS/TTF available on demand. Automated detection reliably identified valid regions across anatomical sites, including head-and-neck scans, a near-worst-case test of uniform-region availability. Both NPS and TTF pipelines were validated against phantom-derived references, confirming that in vivo measurements reflect genuine anatomical texture and resolution rather than pipeline artifacts. By eliminating operator variability and phantom requirements, AnatomIQ enables practical continuous quality monitoring and evidence-based reconstruction optimization. The toolkit is open-source, with optional commercial licensing for academic and clinical use.

## 1. Methods

Contrast-to-noise ratio (CNR) assessment for CT image quality traditionally requires manual selection of lesion and background regions of interest, introducing substantial operator-dependent variability and requiring considerable time for typical research studies. We developed AnatomIQ, an open-source Python toolkit providing automated background detection, PACS integration, and dose-optimization metrics.

### 1.1 System Architecture

AnatomIQ is implemented in Python 3.8+ using standard scientific libraries (NumPy, SciPy, scikit-image) and medical imaging tools (PyDICOM). The modular architecture consists of five core components:

1. **DICOM Handler** - Loads CT images with Hounsfield unit conversion and metadata extraction
2. **Background Detector** - Automated background region identification
3. **Quality Metrics** - Comprehensive image quality calculations including CNR, SNR, dose metrics, NPS, and TTF
4. **Visualization** - Publication-quality figure generation
5. **Web Interface** - Gradio-based GUI for clinical workflow integration

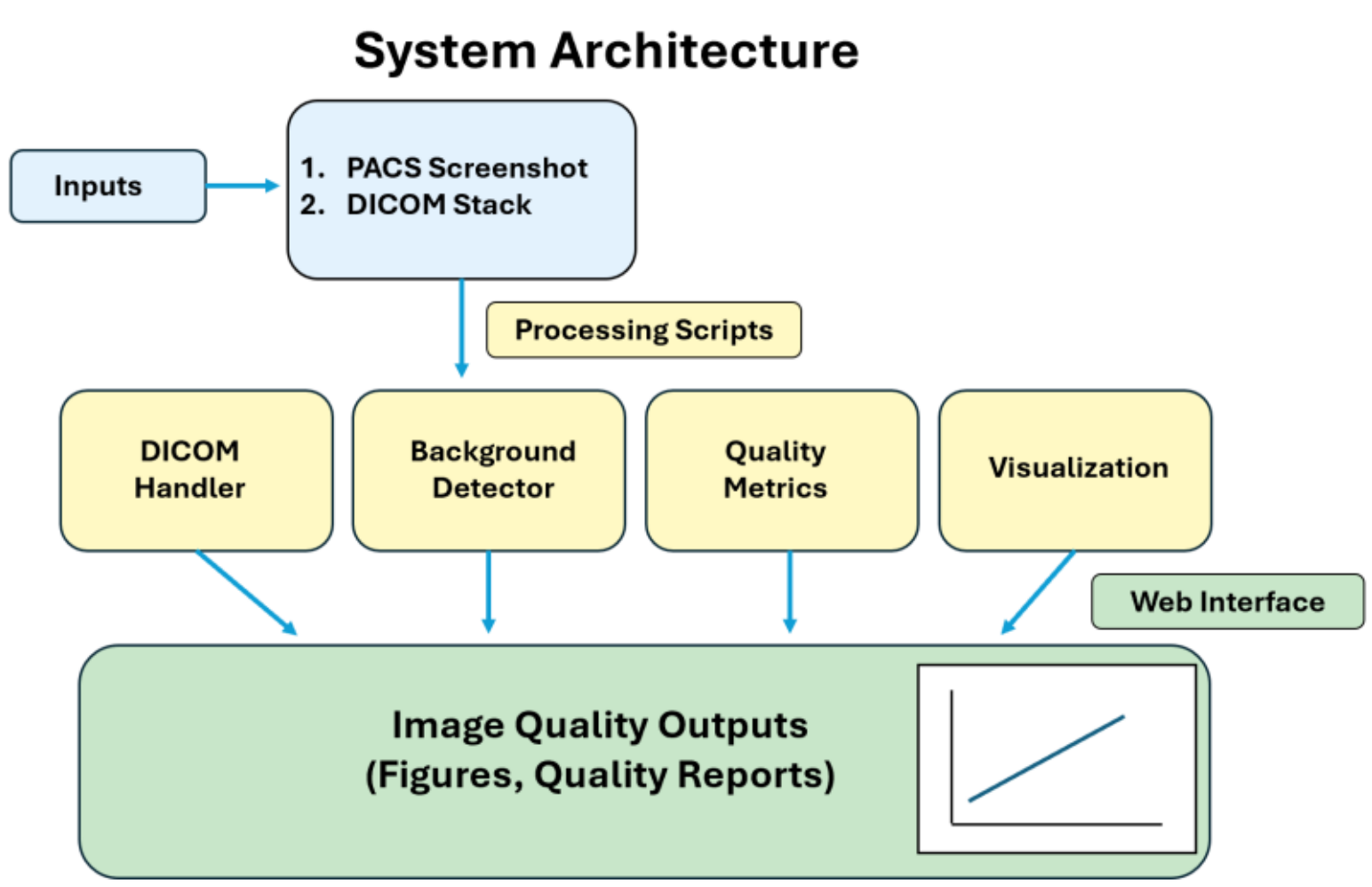

**Figure 1:** System architecture of AnatomIQ showing modular design with data flow from PACS Screenshot and DICOM inputs through processing modules to comprehensive quality outputs. The modular structure enables custom workflows, batch processing, and easy extension.

### 1.2 Automated Background Detection Algorithm

The background detection algorithm addresses three key challenges: tissue specificity, spatial validity, and region homogeneity.

#### 1.2.1 Tissue-Specific HU Thresholding

Rather than assuming subcutaneous fat as background, the algorithm adaptively selects background tissue based on the lesion's HU characteristics. For fat tissue (−150 to −30 HU), the background HU range is set to [−190, −30][1,2]. For soft tissue lesions (−30 to 60 HU)[3], the background HU range is defined as lesion_mean ± 30 HU. For enhanced or dense tissue (60 to 200 HU)[4], the background HU range is set to lesion_mean ± 40 HU. This adaptive approach ensures appropriate tissue matching across different anatomical regions and lesion types.

**1.2.2 Spatial Exclusion Zones**

To prevent partial volume effects and maintain local tissue context, the algorithm implements dual spatial constraints that are fully adjustable to accommodate lesions of varying sizes and clinical scenarios (Figure 2). The inner exclusion zone establishes a minimum distance from the lesion ROI edge to prevent signal spillover due to partial volume effects. The default settings provide a 5 mm user-adjustable buffer within the lesion boundary via an interactive slider to mimic the lesion contour and morphology. The outer limit zone defines the maximum distance from the lesion to ensure background noise characteristics remain comparable to the lesion region, automatically adjusted to maintain optimal spatial relationships. This flexible spatial constraint system enables accurate boundary detection across lesion sizes from small nodules (~5 mm) to large masses (20+ mm). Distance transforms compute pixel-wise distances from the lesion mask, enabling precise spatial filtering with sub-pixel accuracy (i.e., <0.5 mm).

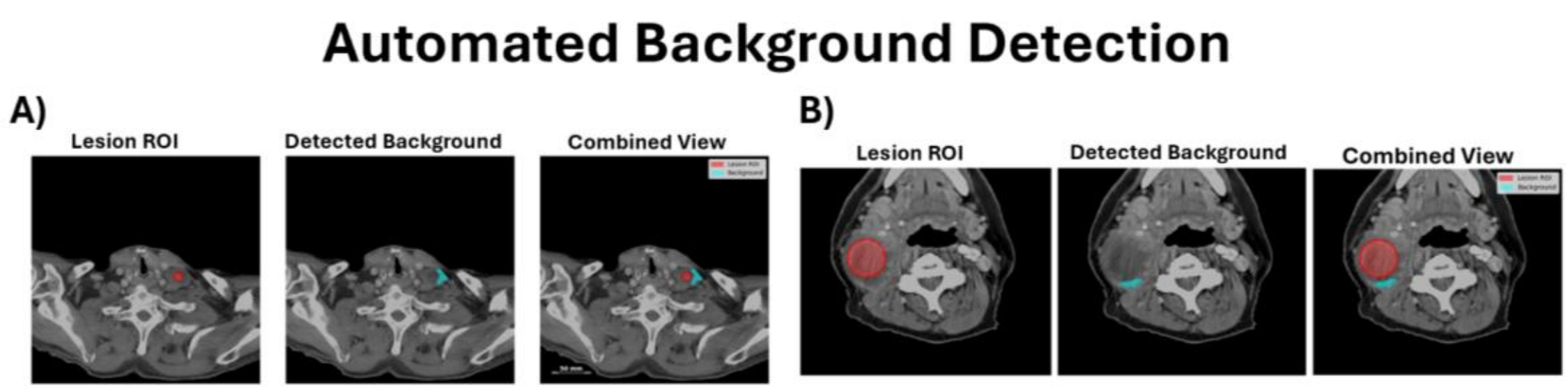

**Figure 2A-B:** Automated background detection with adjustable spatial constraints for lesions of varying sizes. (A) Small lesion with 5 mm inner exclusion zone showing lesion ROI (red) and detected background region (cyan) with appropriate spatial separation. (B) Large lesion with 16 mm inner exclusion zone demonstrating algorithm adaptability for larger masses requiring expanded exclusion to prevent partial volume contamination. The outer limit zone automatically adjusts based on inner exclusion settings to maintain optimal local tissue context.

**1.2.3 Morphological Filtering**

Connected component analysis with morphological operations removes unsuitable regions from background ROI candidates, thereby ensuring measurement stability. The algorithm first removes connected regions smaller than 25% of the minimum required area (default: 100 $mm^2$), eliminating small anatomical structures (vessels, trabecular bone) and imaging artifacts (streak artifacts, metallic artifacts from surgical hardware) that would compromise noise homogeneity. Small holes within remaining regions (< 20 pixels) are then filled to address partial volume effects, air-tissue interfaces, and reconstruction artifacts. Finally, binary opening with a 2-pixel disk structuring element removes thin linear structures (fascia, vessel walls) and edge artifacts that increase spatial noise variance. The algorithm selects the most homogeneous region (i.e., lowest coefficient of variation) that meets the size criteria, ensuring reliable noise statistics.

### 1.2.4 Quality Assessment

Each background region is evaluated using three metrics: area adequacy (≥100 $mm^2$ required for reliable statistics)[5], homogeneity (coefficient of variation < 0.3 preferred)[6], and an overall quality flag classification (GOOD, WARNING, or FAILED) derived from these thresholds.

## 1.3 PACS Integration Workflow

The user captures the PACS display, defines the DICOM region (one-time setup), clicks on the lesion for automatic coordinate extraction, navigates to the matching DICOM slice, and confirms registration via visual overlay. This workflow eliminates manual coordinate transcription and associated errors.

## 1.4 Quality Metrics and Dose Optimization

AnatomIQ calculates comprehensive metrics aligned with medical physics standards. Core metrics include CNR ($|HU_{lesion} - HU_{background}|/\sigma_{background}$), SNR ($HU_{mean}/\sigma$), and Rose detectability index[7,8] ($CNR \times \sqrt{Area_{lesion}}$). Dose-normalized metrics include FOM[9,10] ($CNR^2/SSDE$) and dose efficiency ($CNR/\sqrt{SSDE}$).

The toolkit predicts CNR at alternative dose levels using the noise-dose relationship where noise ∝ $1/\sqrt{dose}$, enabling automated recommendations for protocol optimization. The system identifies adequate contrast quality (CNR ≥ 5.0), marginal contrast quality (CNR 3.0-4.9), and inadequate contrast quality (CNR < 3.0) based on the Rose criterion for lesion detectability[7,8]providing specific image quality adjustment factors.

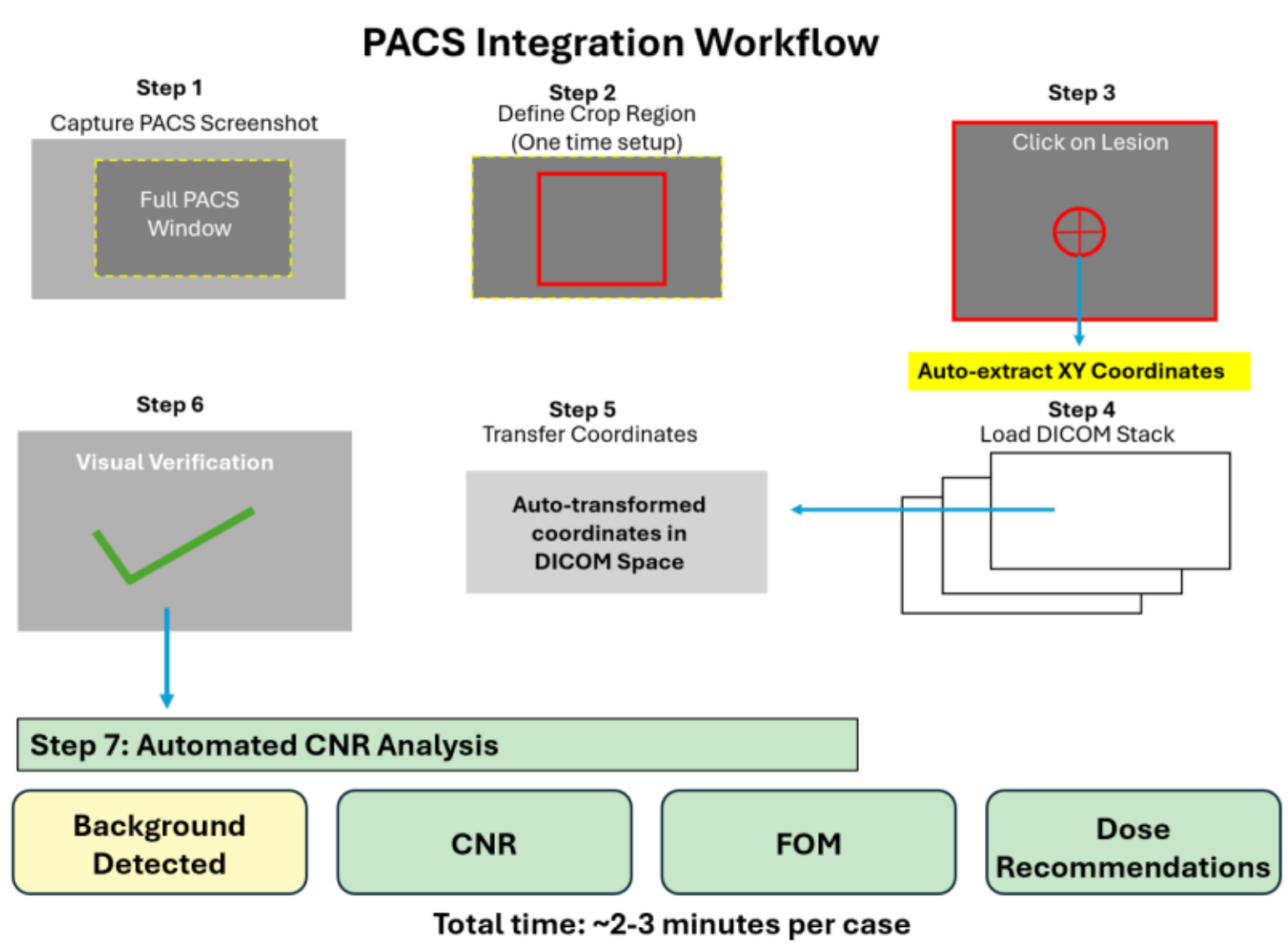

**Figure 3:** PACS integration workflow demonstrating the click-to-select process. Step 1: Capture PACS screenshot. Step 2: Define crop region (one-time setup). Step 3: Click on lesion for automatic coordinate extraction. Step 4: Load DICOM stack. Step 5: Transfer coordinates automatically. Step 6: Visual verification. Step 7: Automated CNR analysis with quality metrics. Total analysis time: 2-3 minutes per case.

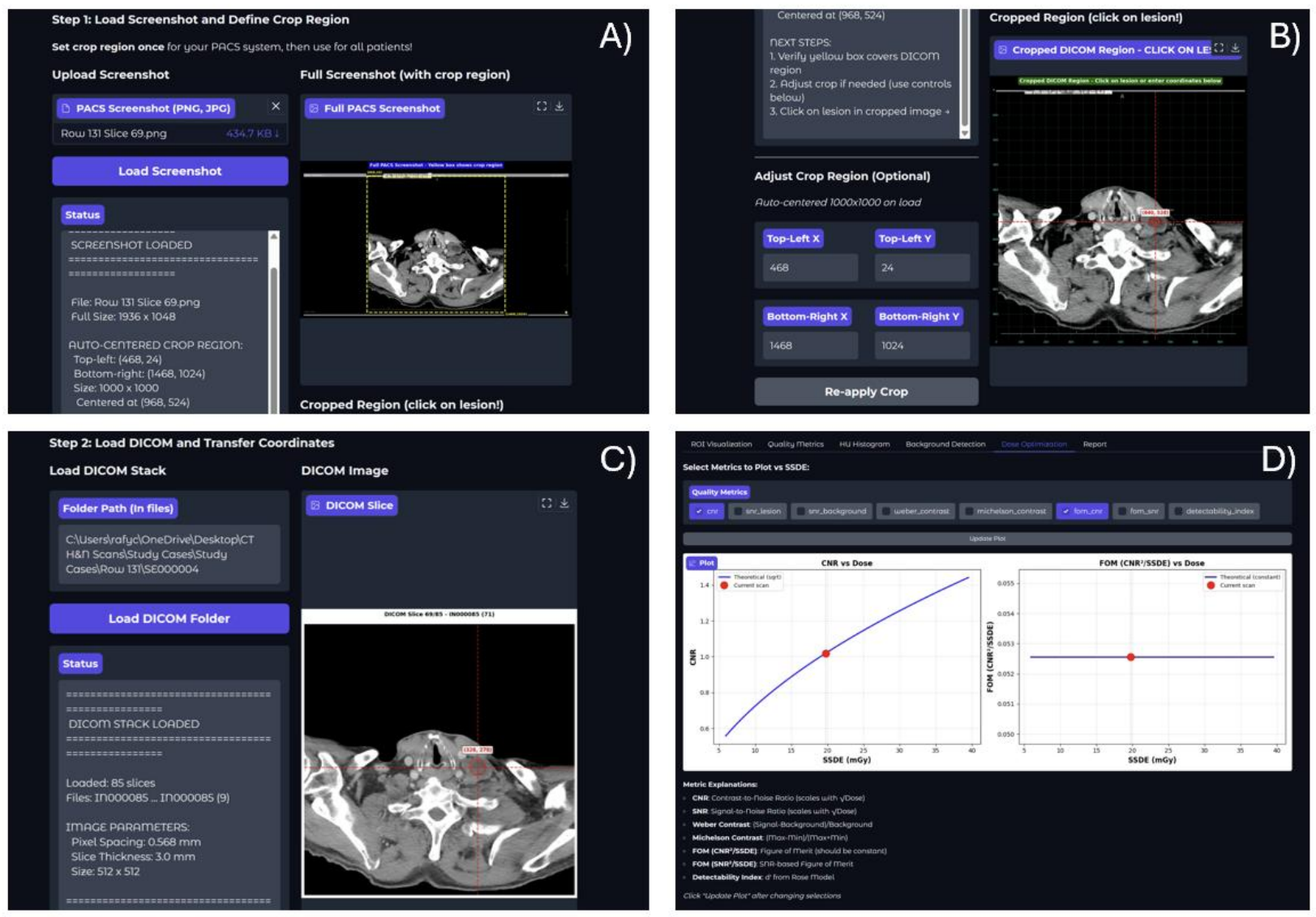

**Figure 4A-D:** PACS integration workflow for automated CNR analysis. (A) Screenshot upload with one-time DICOM region cropping to define the image boundaries. (B) Lesion selection interface displaying PACS annotations with automatic coordinate extraction from user clicks. (C) Coordinate mapping to DICOM slice space with visual verification overlay. (D) Interactive dose-quality analysis with user-selectable metrics plotted against SSDE for protocol optimization.

### 1.5 Advanced Spatial Frequency Analysis

Building on the automated background detection feature, AnatomIQ implements two advanced image quality metrics that characterize fundamental imaging physics trade-offs in Noise Power Spectrum (NPS) for noise characterization and Task Transfer Function (TTF) for spatial resolution assessment. Critically, both metrics are calculated directly from clinical patient scans using anatomical structures as natural measurement tools, eliminating the workflow disruption and radiation exposure of dedicated phantom acquisitions.

#### 1.5.1 Noise Power Spectrum

The NPS quantifies noise magnitude and texture in the spatial frequency domain, providing insight into reconstruction algorithm behavior that simple standard deviation measurements cannot capture[11,12]. Because non-linear iterative and deep-learning reconstruction (IR/DLIR) algorithms introduce spatial noise correlations that extend beyond the patch sizes adequate for filtered back-projection, the algorithm builds a multi-region ensemble consistent with AAPM TG-233 guidance for ensemble-averaged periodogram estimation[5,13]. From the CNR background mask, the algorithm exhaustively searches all valid 32x32-pixel candidate patches within soft tissue, ranks them by variance, and retains the M lowest-variance regions (default M = 16). This automatically excludes patches that intersect fascial boundaries, vessel walls, or bone margins[14]. To accommodate anatomical sites where larger homogeneous regions are available, the patch search begins at 256x256 pixels and automatically downscales (256 -> 128 -> 64 -> 32 px) only when an insufficient number of valid low-variance candidates are found at the current scale. This reactive sizing was validated using a uniform CT phantom, where the algorithm consistently selected the full 256x256 px patch size given the phantom's large homogeneous field, in contrast to patient head-and-neck scans, which consistently required downscaling to 32×32 px. NPS candidate patches were drawn from within the same HU-thresholded, morphologically filtered homogeneous-tissue mask used for CNR background detection (Section 1.2). The M = 16 lowest-variance windows were then selected from this mask using exhaustive patch-level search, without manual placement. Each patch undergoes second-order polynomial surface detrending (fit and subtraction of a quadratic surface, $ax^2 + by^2 + cxy + dx + ey + f$) to remove curved tissue gradients and cupping artifacts. These artifacts would otherwise leak into the Fourier domain, so they are tapered with a 2D Hann window and zero-padded from 32×32 to 64×64 pixels before the 2D FFT. Zero-padding performs smooth sinc interpolation across the frequency grid without altering integrated noise energy, eliminating the empty-bin ("picket-fence") artifacts that arise from small patch sizes. Following the classical ensemble-averaged NPS formulation[15], window power is corrected using energy-preserving normalization to account for the Hann taper applied to each patch:

$$NPS = \frac{|FFT_2D|^2 \cdot \Delta x \cdot \Delta y}{(N^2 \cdot C_{win})} \quad (1)$$

where $C_{win} = \left(\frac{1}{N^2}\right)\Sigma W^2$ ensuring compliance with Parseval's theorem ($\int NPS(f)df \approx \sigma^2$), reducing to the standard rectangular-window NPS estimator[15] when $C_{win} = 1$ (no apodization). A smooth physical rolloff, $1 - e$^$(-(f/0.045)^2)$, replaces a hard low-frequency cutoff to model the zero-gain-at-DC behavior of the CT reconstruction ramp filter while avoiding artificial step-edge ringing. The M single-patch estimates are radially averaged into the reported ensemble 1D and 2D NPS curves, in units of $HU^2 \cdot mm^2$.

Key metrics include average NPS magnitude (overall noise power), peak spatial frequency (dominant noise texture: fine-grained vs. coarse), noise texture classification (FINE/MODERATE/COARSE based on peak frequency), and dose-normalized NPS. Consistent with AAPM TG-233 and ICRU recommendations for ensemble periodogram averaging (M ≥ 8–16 independent regions)[5,16], the analysis is flagged as low-confidence when fewer than eight qualifying low-variance patches are available within the uniform tissue mask which corresponds to a scenario more common in anatomically constrained regions such as the head and neck.

**1.5.2 Task Transfer Function**

TTF quantifies spatial resolution by measuring the system's ability to reproduce contrast across different spatial frequencies[17]. Unlike traditional phantom-based TTF measurements, AnatomIQ leverages natural anatomical edges in patient scans. The algorithm automatically detects circular structures (trachea, blood vessels, airways) using Hough circle detection[18] with HU-based contrast thresholding (>200 HU contrast), extracts 180 radial intensity profiles across the circular edge, averages profiles to generate the Edge Spread

Function (ESF) representing the intensity transition, differentiates ESF to obtain Line Spread Function (LSF), Fourier transforms LSF to produce TTF curve, and calculates resolution metrics.

Key outputs include TTF50 (i.e., spatial frequency at 50% modulation, reported in $mm^{-1}$), TTF10 (i.e., frequency at 10% modulation), spatial resolution in line pairs per millimeter (lp/mm), TTF at Nyquist frequency, resolution category (i.e., EXCELLENT: >0.55 $mm^{-1}$; GOOD: 0.45-0.55; FAIR: 0.35-0.45; POOR: <0.35), and structure quality score (i.e., 0-1 scale assessing edge sharpness and circularity). The circular-edge detection and TTF calculation pipeline was additionally validated on a resolution phantom containing high-contrast, precisely circular density inserts, providing a known-geometry edge against which the automated ESF extraction, LSF differentiation, and TTF derivation could be verified independent of anatomical variability.

The anatomical TTF approach using *in vivo* structures provides clinically relevant resolution measurements that reflect actual imaging conditions such as patient motion, positioning, and physiological factors rather than idealized phantom scenarios. These features are implemented as on-demand calculations to maintain fast workflow for routine analysis.

**2. Results**

AnatomIQ generates comprehensive output visualizations including ROI placement, HU distribution histograms, and complete image quality metrics (CNR, SNR, detectability indices, SSDE-normalized FOM) (Figure 5), phantom-validated, phantom-free noise power spectrum analysis with texture classification (Figure 6), and phantom-validated TTF curves derived from anatomical structures (Figure 7). Automated background detection successfully identified valid regions meeting size (≥100 $mm^2$) and homogeneity (CV < 0.3) criteria across diverse anatomical sites. Analysis time was reduced to 2-3 minutes per case (50% faster than manual ROI placement), with optional on-demand NPS (5-15s) and TTF (10-25s) calculations available when spatial frequency characterization is preferred. The automated approach eliminates inter-operator variability while maintaining consistent HU-based tissue matching across cases.

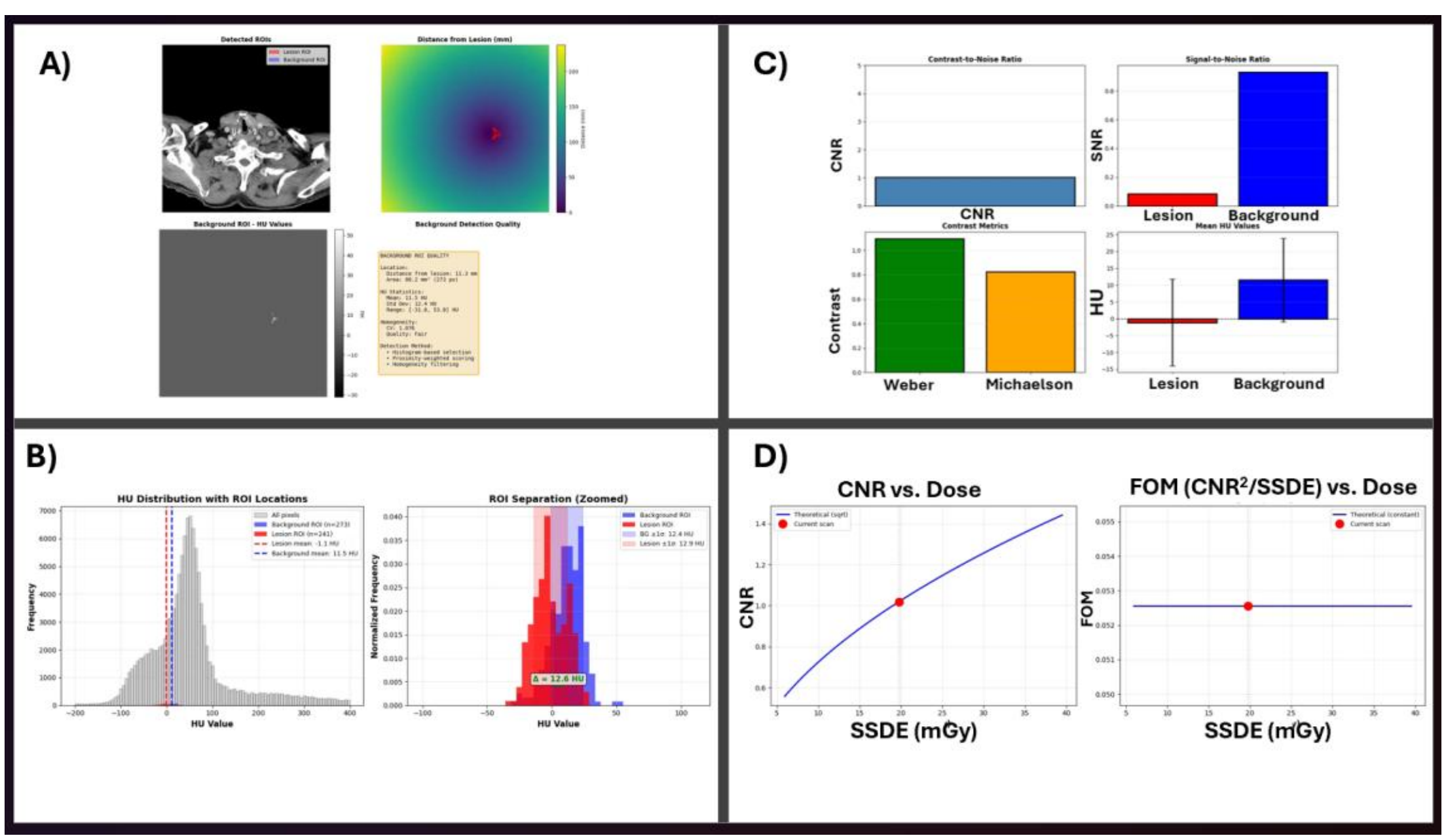

**Figure 5A-D:** Automated background detection and quality assessment outputs. (A) Spatial exclusion zones and distance transform visualization showing detected ROIs, lesion mask, and background region with HU values. (B) Histogram analysis of HU distribution for lesion (red) and background (blue) regions with Gaussian fitting. (C) Comprehensive image quality metrics including CNR, SNR, detectability indices, and QA assessment. (D) Dose-normalized performance plots showing CNR and FOM relationships with SSDE.

Phantom-free NPS analysis was validated using a two-stage approach (Figure 6A-F). Applied to a uniform CT phantom, the algorithm automatically selected the largest candidate patch size (M = 16 ROIs at 256x256 px) given the phantom's large homogeneous field, producing a single, symmetric spectral peak at 0.098 $mm^{-1}$ (average NPS = 58.5 $HU^2 \cdot mm^2$) with a radially isotropic 2D NPS confirming the pipeline does not introduce directional or spectral artifacts under ideal conditions. Applied to a head-and-neck patient scan which we identify as a near-worst-case test of anatomical uniform-region availability, the algorithm automatically downscaled to 32×32 px (M = 16 ROIs), yielding a peak frequency of 0.066 $mm^{-1}$ (average NPS = 646.9 $HU^2 \cdot mm^2$) with an anisotropic 2D NPS oriented along the underlying tissue plane, consistent with genuine anatomical texture rather than measurement artifact. Both results were classified as coarse/smooth noise texture, consistent with IR/DLIR reconstruction.

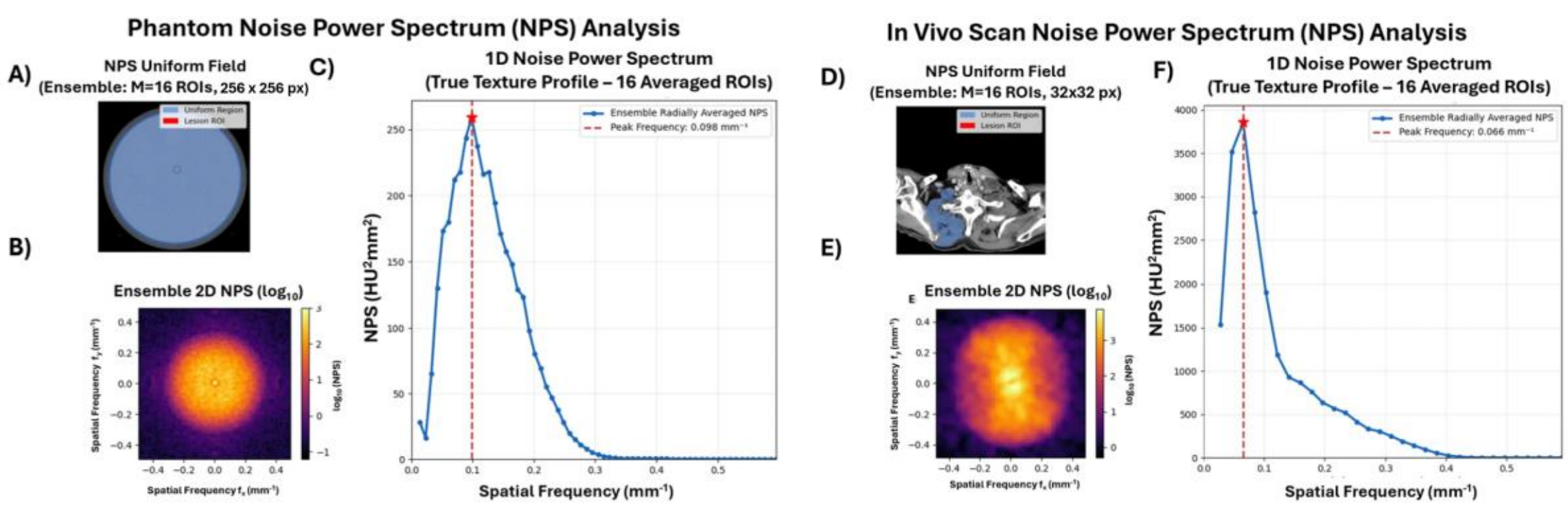

**Figure 6A-F:** Validation of the automated ensemble NPS pipeline on phantom versus *in vivo* data. (A-C) Uniform CT phantom: (A) M = 16 automatically selected 256×256 px regions, (B) isotropic ensemble 2D NPS, and (C) 1D NPS with peak at 0.098 $mm^{-1}$, confirming the pipeline introduces no directional or spectral artifacts under ideal conditions. (D-F) Patient head-and-neck scan: (D) M = 16 regions automatically downscaled to 32×32 px given the anatomically limited uniform field, (E) anisotropic ensemble 2D NPS reflecting genuine tissue texture, and (F) 1D NPS with peak at 0.066 $mm^{-1}$. The phantom's isotropic result serves as an internal control for the *in vivo* measurement.

TTF analysis was validated using the same two-stage phantom/in vivo approach as the NPS pipeline (Figure 7A-D). Applied to a resolution phantom, the algorithm automatically detected a high-contrast circular density insert and produced a clean, well-managed TTF curve (TTF50 = 0.156 $mm^{-1}$, TTF10 = 0.277 $mm^{-1}$), confirming correct edge-spread-function extraction and Fourier-domain TTF derivation on a geometrically ideal, manufactured edge. Applied to the patient scan, the algorithm automatically detected the trachea (structure quality score 0.89, edge contrast 988 HU) as a natural anatomical edge, yielding TTF50 = 0.172 $mm^{-1}$ and TTF10 = 0.298 $mm^{-1}$, below the Nyquist frequency of 0.50 $mm^{-1}$ and indicating a smooth reconstruction kernel optimized for soft-tissue contrast. These *in vivo* values were obtained entirely from patient anatomy without a dedicated phantom acquisition; the phantom result serves only as an internal validation of the underlying pipeline.

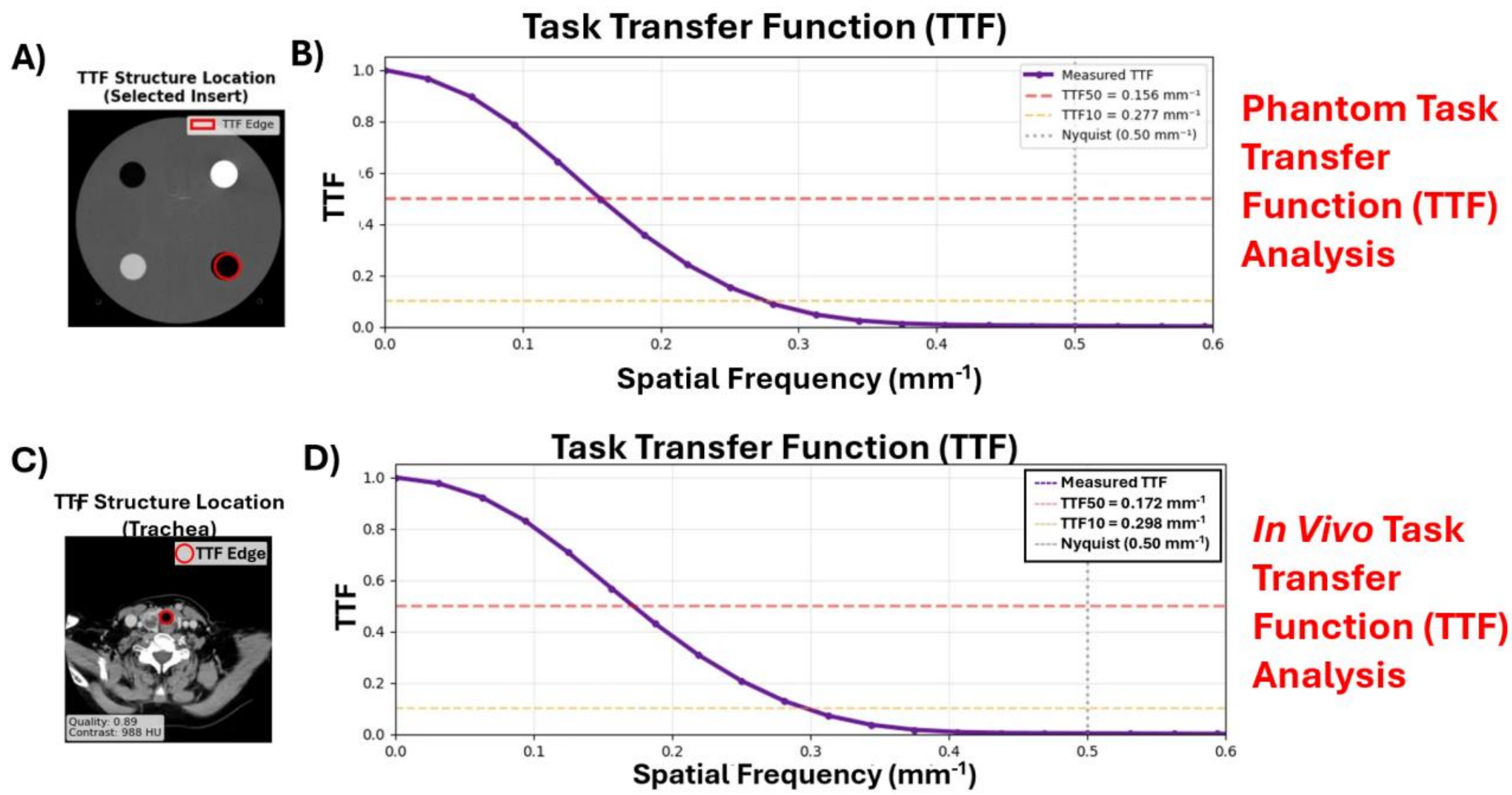

**Figure 7A-D:** Validation of the automated TTF pipeline on phantom versus *in vivo* data. (A-B) Resolution phantom: (A) automatically detected high-contrast circular density insert selected for edge sharpness and circularity, and (B) resulting TTF curve (TTF50 = 0.156 $\text{mm}^{-1}$, TTF10 = 0.277 $\text{mm}^{-1}$), confirming correct edge-detection and Fourier-domain TTF derivation on a geometrically ideal, manufactured edge. (C-D) Patient head-and-neck scan: (C) automatically detected trachea (quality score 0.89, contrast 988 HU) used as a natural anatomical edge, and (D) resulting TTF curve (TTF50 = 0.172 $\text{mm}^{-1}$, TTF10 = 0.298 $\text{mm}^{-1}$), indicating a smooth reconstruction kernel optimized for soft-tissue contrast. The phantom result serves as an internal control confirming the pipeline recovers a physically plausible TTF from a known edge before being applied to anatomical structures.

## 3. Discussion

AnatomIQ addresses critical workflow barriers in CT image quality assessment through automated background detection and PACS integration. Unlike many commercial software packages that lack automated region selection capabilities, AnatomIQ's tissue-adaptive algorithm ensures consistent HU-based tissue matching without operator dependence. The integration of NPS and TTF analysis represents a significant advance: these physics-based metrics are traditionally measured using standardized phantoms[19] requiring separate acquisitions, additional radiation exposure, and workflow interruption. AnatomIQ's approach leverages natural anatomical structures (vessels, airways, trachea) present in clinical scans, providing resolution and noise characterization without additional imaging. This "phantom-free" methodology yields clinically relevant measurements that incorporate real-world factors (patient motion, positioning variability, physiological effects) rather than idealized phantom conditions. The trade-off between spatial resolution (TTF) and noise (NPS) is automatically quantified, enabling objective assessment of reconstruction kernel selection such as when smooth kernels show low TTF50 (<0.25 $mm^{-1}$) but also low NPS, while sharp edge-enhancing kernels show high TTF50 (>0.50 $mm^{-1}$) with correspondingly higher noise. This comprehensive characterization supports evidence-based protocol optimization beyond simple CNR measurements. The elimination of manual background selection enables practical implementation of continuous quality monitoring in clinical settings.

This *in vivo* emphasis also serves to address a structural limitation of phantom-based physics characterization. Standard phantoms (e.g., ACR CT phantom and the Catphan) are homogeneous, static, and geometrically idealized; they cannot reproduce the heterogeneous tissue textures, respiratory and cardiac motion, vascular pulsation, and patient-specific attenuation profiles that shape the noise and resolution a radiologist encounters at the console. Recent work has begun to formalize this gap: automated in vivo NPS methods have been developed and validated against phantom-derived references[20,21], and multi-patient clinical data have shown that individual peak noise magnitude can vary substantially between patients scanned under nominally identical protocols indicating variability where a one-size-fits-all phantom measurement cannot capture[22]. Because non-linear IR and DLIR algorithms apply anatomy and structure-dependent regularization, their noise and resolution behavior on a uniform phantom does not necessarily generalize to real tissue[23]. AnatomIQ's patient-derived NPS and TTF measurements are therefore best understood not as a substitute for phantom-based quality assurance, but as the clinically relevant ground truth against which phantom-based protocols should ultimately be validated: they characterize image quality exactly as it is delivered to the interpreting radiologist, under actual scan conditions, in actual patients.

The pipeline's reactive patch-size selection (256 ->128 ->64 ->32 px) was further validated on a uniform CT phantom, which consistently retained the full patch size given its large homogeneous field, while identical processing of patient head-and-neck scans consistently required downscaling to 32×32 px. This confirms that the ensemble/detrending safeguards developed for constrained anatomy are invoked only when anatomically necessary rather than applied uniformly, and that the pipeline recovers a clean, isotropic NPS under ideal (phantom) conditions before being tested under the anatomically worst-case conditions reported above. One caveat is worth noting when comparing peak frequencies across patch sizes: frequency resolution is set by patch width ($\Delta f \approx 1/L$)[14,24], so the phantom's 256×256 px estimate ($\Delta f \approx 0.0039$ $mm^{-1}$) is resolved substantially more finely than the in vivo 32×32 px estimate ($\Delta f \approx 0.031$ $mm^{-1}$, prior to zero-pad interpolation). Reported peak frequencies at different patch sizes should therefore be interpreted as approximate to their respective resolution floors rather than directly comparable at face value.

The TTF pipeline was validated analogously, using a high-contrast circular density insert in a resolution phantom to confirm correct edge-spread-function extraction and Fourier-domain TTF derivation (TTF50 = 0.156 $mm^{-1}$) independent of the physiologic edge irregularity present in anatomical structures. Because the phantom insert differs from the *in vivo* trachea in both material composition and the reconstruction protocol under which it was scanned, its TTF50 is not expected to numerically match the *in vivo* result (0.172 $mm^{-1}$); rather, the phantom measurement serves to confirm that the algorithm correctly recovers a physically plausible TTF curve from a geometrically ideal edge, providing confidence that the same pipeline applied to anatomical edges is measuring genuine resolution rather than an artifact of the detection or curve-fitting process.

Preliminary validation was intentionally conducted on head-and-neck CT scans, representing a near-worst-case scenario for phantom-free spatial frequency analysis: the small field of view, complex multi-tissue anatomy, and scarcity of large homogeneous regions make this among the most demanding anatomical sites for automated uniform-region detection. As reported above (Figures 6-7), the automated methods successfully identified suitable structures for both NPS and TTF measurement despite these constraints, demonstrating robust performance under anatomically challenging conditions and, by extension, suggesting that larger and more homogeneous anatomical regions should perform at least as well with less computational overhead. The roughly order-of-magnitude difference in NPS amplitude between the phantom (Figure 6C, peak ≈ 260 $HU^2 \cdot mm^2$) and *in vivo* (Figure 6F, peak ≈ 3800 $HU^2 \cdot mm^2$) measurements is not attributable to dose, as the patient scan was acquired at a higher SSDE than the phantom. Rather, it reflects the fundamental difference between a spectrally flat, homogeneous phantom material and real soft tissue, whose microstructural texture is not fully removed by second-order polynomial detrending and is captured by the Fourier transform as additional noise power[23]. This is consistent with the anisotropic 2D NPS observed *in vivo* (Figure 6E) versus the isotropic phantom result (Figure 6B) and further supports the position that phantom-derived noise metrics systematically underestimate the noise magnitude presented to the interpreting radiologist.

The 32×32 px ensemble configuration used here was itself a consequence of this worst-case anatomy: fixed default patch sizes intended to capture the longer-range spatial noise correlations characteristic of IR/DLIR reconstruction repeatedly failed to locate valid regions in the neck, where no homogeneous soft-tissue area of that width exists, triggering fallback patches that spanned bone-airway-muscle boundaries. This motivates an anatomy-aware, case-based patch-sizing strategy as a direction for future development where sites with larger native uniform areas; i.e. abdominal scans where solid organs such as the liver readily provide multi-centimeter homogeneous fields, should tolerate larger default patch sizes without requiring the zero-padding and variance-ranking safeguards developed here for anatomically constrained sites, reducing both computation time and the number of ROIs needed for a stable ensemble estimate. Planned future work includes automatic patch-size selection driven by the segmented area of the detected uniform-tissue mask: applying the constrained-anatomy pipeline (32×32 px, M = 16 ensemble, polynomial detrending) only where the uniform region falls below a minimum area threshold, and reverting to more direct, larger-patch sampling where larger organs (liver, spleen, descending aorta) permit it. This is expected to further improve measurement reliability, reduce computation time, and expand the method's applicability across anatomical sites without sacrificing the robustness demonstrated here in the most anatomically constrained case.

The dose-normalized metrics (FOM, dose efficiency) enable direct protocol comparison across different scanners and patient sizes, supporting adaptive imaging protocols. The system identifies adequate contrast quality (CNR ≥ 5.0), marginal contrast quality (CNR 3.0-4.9), and inadequate contrast quality (CNR < 3.0) scenarios based on validated observer detection thresholds[7,8]providing specific dose adjustment recommendations for follow-up scans. Current CNR and SNR acceptability thresholds are applied uniformly across anatomical sites; however, published detectability thresholds vary meaningfully by body region and clinical task[25,26]. Future development will incorporate automated anatomical-region identification to apply region-appropriate thresholds accompanying the anatomy-aware patch-sizing strategy described above for NPS.

## 4. Conclusion

AnatomIQ provides automated, reproducible CNR analysis for CT image quality assessment, eliminating operator variability while reducing analysis time by 50%. The open-source toolkit enables practical quality monitoring and dose optimization through seamless PACS integration and comprehensive image quality metrology. The modular design facilitates community enhancement and clinical integration. Uniquely, the toolkit performs advanced NPS and TTF analysis directly on patient anatomy, providing comprehensive resolution and noise characterization that reflects real anatomy, motion, and dose conditions rather than idealized phantom surrogates. This enables practical, routine assessment of imaging physics, grounded in the same conditions under which images are clinically interpreted, that was previously limited to research settings with dedicated phantom protocols.

## References

1. Kvist H, Chowdhury B, Grangård U, Tylén U, Sjöström L. Total and visceral adipose-tissue volumes derived from measurements with computed tomography in adult men and women: predictive equations. *Am J Clin Nutr.* 1988;48(6):1351-1361.
2. Yoshizumi T, Nakamura T, Yamane M, et al. Abdominal fat: standardized technique for measurement at CT. *Radiology.* 1999;211(1):283-286.
3. Lamba R, McGahan JP, Corwin MT, et al. CT Hounsfield numbers of soft tissues on unenhanced abdominal CT scans: variability between two different manufacturers' MDCT scanners. *AJR Am J Roentgenol.* 2014;203(5):W474-W479.
4. Bae KT. Intravenous contrast medium administration and scan timing at CT: considerations and approaches. Radiology. 2010;256(1):32-61.
5. Samei E, et al. Performance evaluation of computed tomography systems: Summary of AAPM Task Group 233. Med Phys. 2019;46(11):e735-e756.
6. Christianson O, Winslow J, Frush DP, Samei E. Automated technique to measure noise in clinical CT examinations. *AJR Am J Roentgenol.* 2015;205(1):W93-W99.
7. Rose A. The sensitivity performance of the human eye on an absolute scale. J Opt Soc Am. 1948;38(2):196-208.
8. Christianson O, Chen JJ, Yang Z, et al. A minimum SNR criterion for computed tomography object detection in the projection domain. Med Phys. 2022;49(8):5107-5121.
9. American Association of Physicists in Medicine. Size-Specific Dose Estimates (SSDE) in Pediatric and Adult Body CT Examinations. AAPM Report No. 204. College Park, MD: AAPM; 2011.
10. Kawashima H, Ichikawa K, Hanaoka S, Matsubara K, Takata T. Relationship between size-specific dose estimates and image quality in computed tomography depending on patient size. J Appl Clin Med Phys. 2018 Jul;19(4):246-251. doi:10.1002/acm2.12340. Epub 2018 May 4. PMID: 29729075; PMCID: PMC6036365.

11. Friedman SN, Fung GS, Siewerdsen JH, Tsui BM. A simple approach to measure computed tomography (CT) modulation transfer function (MTF) and noise-power spectrum (NPS) using the American College of Radiology (ACR) accreditation phantom. Med Phys. 2013;40(5):051907. doi:10.1118/1.4800795.

12. Boedeker KL, Cooper VN, McNitt-Gray MF. Application of the noise power spectrum in modern diagnostic MDCT: part I. Measurement of noise power spectra and noise equivalent quanta. Phys Med Biol. 2007 Jul 21;52(14):4027-46. doi:10.1088/0031-9155/52/14/002. Epub 2007 Jun 8. PMID: 17664593.

13. Welch PD. The use of fast Fourier transform for the estimation of power spectra. IEEE Trans Audio Electroacoust. 1967;15(2):70-73.

14. Dolly S, Chen HC, Anastasio M, Mutic S, Li H. Practical considerations for noise power spectra estimation for clinical CT scanners. J Appl Clin Med Phys. 2016;17(3):350-365. doi:10.1120/jacmp.v17i3.5841

15. Bushberg JT, Seibert JA, Leidholdt EM Jr, Boone JM. The Essential Physics of Medical Imaging. 4th ed. Philadelphia, PA: Wolters Kluwer; 2021.

16. Pahn G, Skornitzke S, Schlemmer HP, Kauczor HU, Stiller W. Toward standardized quantitative image quality (IQ) assessment in computed tomography (CT): A comprehensive framework for automated and comparative IQ analysis based on ICRU Report 87. *Physica Medica.* 2016;32(1):104-115. doi:10.1016/j.ejmp.2015.09.017

17. Richard S, Husarik DB, Yadava G, Murphy SN, Samei E. Towards task-based assessment of CT performance: System and object MTF across different reconstruction algorithms. Med Phys. 2012;39(7):4115-4122. doi:10.1118/1.4725171.

18. Duda RO, Hart PE. Use of the Hough transformation to detect lines and curves in pictures. Commun ACM. 1972;15(1):11-15.

19. Chun M, Choi JH, Kim S, Ahn C, Kim JH. Fully automated image quality evaluation on patient CT: Multi-vendor and multi-reconstruction study. PLoS One. 2022 Jul 20;17(7):e0271724. doi: 10.1371/journal.pone.0271724. PMID: 35857804; PMCID: PMC9299323.

20. Sanders J, Hurwitz L, Samei E. Patient-specific quantification of image quality: An automated method for measuring spatial resolution in clinical CT images. Med Phys. 2016 Oct;43(10):5330. doi:10.1118/1.4961984. PMID: 27782718.

21. Smith TB, Abadi E, Sauer TJ, Fu W, Solomon J, Samei E. In vivo noise texture estimation: Development and validation of an automated methodology. Proc SPIE Prog Biomed Opt Imaging. 2020.

22. Park C, Kim S, Kim JH. Noise power spectrum analysis in CT for improved patient-specific image optimization: a shift from phantom model to clinical scan. Proc SPIE Med Imaging 2025. 2025;13405:134054F. doi:10.1117/12.3046950

23. Solomon J, Samei E. Quantum noise properties of CT images with anatomical textured backgrounds across reconstruction algorithms: FBP and iterative. Med Phys. 2014;41(9):091908.

24. Bracewell RN. *The Fourier Transform and Its Applications*. 3rd ed. New York, NY: McGraw-Hill; 2000.

25. Ott JG, Ba A, Racine D, et al. Patient-based assessment of quantitative image quality in chest CT: A feasibility study. Eur Radiol. 2022;32:1861-1872. doi:10.1007/s00330-021-08240-4.

26. Verdun FR, Racine D, Ott JG, et al. Image quality in CT: From physical measurements to model observers. Phys Med. 2015;31(8):823-843. doi:10.1016/j.ejmp.2015.08.007.